\documentclass[12pt]{article}
\usepackage{amsfonts, amssymb}
\usepackage[arrow,matrix,curve,cmtip,ps]{xy}
\usepackage{amsthm}
\usepackage{amsmath}
\usepackage[usenames]{color}
\input xy
\xyoption{all}

\newcommand{\maxtensor}{\otimes_{\mbox{max}}}
\newcommand{\mintensor}{\otimes_{\mbox{min}}}

\newcommand{\stensor}{\otimes_{\mbox{str}}}
\newcommand{\face}{\mbox{face}}

\newcommand{\im}{\mbox{im}}
\renewcommand{\1}{\mbox{{\bf 1}}}

\newtheorem{theorem}{Theorem}[section]
\newtheorem{lemma}[theorem]{Lemma}
\newtheorem{proposition}[theorem]{Proposition}
\newtheorem{corollary}[theorem]{Corollary}
\newtheorem{fact}[theorem]{Fact}

\newtheorem{example}[theorem]{Example}

\newtheorem{conjecture}[theorem]{Conjecture}
\newtheorem{definition}[theorem]{Definition}
\newtheorem{problem}{Open Problem}
\theoremstyle{remark}
\newtheorem{remark}{Remark}
\newtheorem{question}{Question}

\newcommand{\tr}{\text{Tr}}

\newcommand{\spn}{\text{span}}
\newcommand{\beq}{\begin{equation}}
\newcommand{\eeq}{\end{equation}}
\newcommand{\beqa}{\begin{eqnarray}}
\newcommand{\eeqa}{\end{eqnarray}}
\renewcommand{\phi}{\varphi}
\def\qed{{\hfill $\blacksquare$}}
\def\proof{\noindent {\bf Proof:  }}

\def\half{{\frac{1}{2}}}
\def\degree{{\circ}}

\def\H{{\mathbb{H}}}
\def\R{{\mathbb{R}}}
\def\C{{\mathbb{C}}}

\def\iso{\simeq}

\def\intersect{\cap}
\def\interior{{\rm int~}}
\def\id{{\mathrm{id}}}
\def\face{{\rm Face}}

\def\ker{{\rm ker~}}
\def\lin{{\mathrm{lin~}}}
\def\homega{\hat{\omega}}

\def\L{{\mathcal L}}

\newcommand{\hbcomment}[1]{{\small \color{blue} [HB: {#1}] \\}}
\newcommand{\hbcommentoff}[1]{{}}
\newcommand{\tempout}[1]{{}}

\title{Ensemble Steering, Weak Self-Duality, and the Structure of Probabilistic Theories}

\author{Howard Barnum\footnote{Perimeter Institute for Theoretical
    Physics; {\tt hbarnum@perimeterinstitute.ca}}, Carl Philipp
  Gaebler\footnote{Harvey Mudd College} and Alexander
  Wilce\footnote{Susquehanna University; {\tt wilce@susqu.edu}}}

\begin{document}
\date{March 4, 2010}

\maketitle

\begin{abstract} 
In any probabilistic theory, we may say a bipartite state $\omega$ on
a composite system $AB$ {\em steers} its marginal state $\omega^{B}$
if, for any decomposition of $\omega^{B}$ as a mixture $\omega^{B} =
\sum_i p_i \beta_i$ of states $\beta_i$ on $B$, there exists an
observable $\{a_i\}$ on $A$ such that the conditional states
$\omega_{B|a_i}$ are exactly the states $\beta_i$.  This is always so
for pure bipartite states in quantum mechanics, a fact first observed
by Schr\"{o}dinger in 1935.  Here, we show that, for {\em weakly}
self-dual state spaces (those isomorphic, but perhaps not canonically
isomorphic, to their dual spaces), the assumption that every state of
a system is steered by some bipartite state on two copies of that system, of a composite  
amounts to the homogeneity of the cone of unnormalized 
states. If the state space is actually self-dual, and not just
weakly so, this implies (via the Koecher-Vinberg Theorem) that it is
the self-adjoint part of a formally real Jordan algebra, and hence,
quite close to being quantum mechanical.

\end{abstract}

\section{Introduction}\label{sec: intro}

The founders of quantum mechanics were already well aware that some of
its most non-classical (and seemingly paradoxical) aspects are
naturally understood in terms information. The Bohrian notion of
complementarity, for example, can be understood in terms of one type
of knowledge about a system precluding another. The fact that
measurement must disturb the state of a system, also stressed by Bohr,
again gives fundamental status to a notion closely connected to
information, the notion of measurement, whose essence is the
acquisition of some sort of information about a system.  Schr\"odinger
was particularly inclined to take this point of view, as evidenced by
his description of entanglement:
\begin{quote}
The best possible knowledge of a total system does not necessarily
include total knowledge of all its parts, not even when these are
fully separated from each other and at the moment are not influencing
each other at all. \cite{Schrodinger}
\end{quote}

The development of quantum information theory has rekindled interest
in the possibility of characterizing quantum theory in operational or
information-theoretic terms. Characteristically, this newer work has
focused on finite-dimensional systems, and has emphasized
considerations involving composite systems. It has become clear that
many properties of quantum systems, e.g., the existence and basic
properties of entangled states, are much better understood as
generically non-classical, rather than specifically quantum,
phenomena, in the sense that they arise in arbitrary non-classical
probabilistic theories \cite{Barrett, BBLW06, BBLW08, Foulis-Randall,
  Klay, KRF, Wilce92}. There is therefore a premium on identifying
operationally meaningful properties of bipartite quantum states that
are, so to say, parochial---that is, properties that are {\em not}
generic in this way. An example is the principle of information
causality, recently introduced in \cite{PPKSWZ09}.  Ideally, one would
like to find a small set of such principles that pick out quantum
mechanics on the nose, but failing this, it is still of interest to
identify principles that define a small neighborhood of theories close
to quantum mechanics.

A property of entangled quantum states that struck Schr\"odinger as especially
odd is the fact that an observer controlling one component of such a
state can {\em steer} the other system into any statistical ensemble
for its (necessarily, mixed) marginal state, simply by choosing to
measure a suitable observable \cite{Schrodinger, HJW}. 
\begin{quote} It is rather discomforting that the [quantum] theory should allow a system to be steered or piloted
into one or the other type of state at the experimenter's mercy in
spite of his having no access to it. \cite{Schrodinger}
\end{quote} 
What Schr\"odinger found discomforting is now understood to be an
important information theoretic feature of quantum mechanics.  This
became clear when Bennett and Brassard \cite{BB84}, in the same paper
that introduced quantum key distribution, considered a natural quantum
scheme for another important cryptographic primitive, bit commitment,
and showed that ensemble steering can be used to break it.\footnote{In
  the scheme, the two possible values Alice can commit to are
  represented by two distinct ensembles for the same density matrix;
  she is to send samples from the ensemble to Bob in order to commit,
  and later reveal which states she drew so that Bob can check that
  she used the claimed ensemble.  However, by sending, not a draw from
  the ensemble but one system of a pure bipartite entangled state with
  the specified density matrix, and keeping the other system, she can
  realize either ensemble after she's already sent the systems to Bob
  by making measurements on her entangled system, enabling her to
  perfectly mimic commitment to either bit.}

In this paper, we connect the possibility of ensemble steering with
two very special geometric properties shared by finite-dimensional
classical and quantum state spaces.  First, such state spaces are {\em
  self-dual}: their cones of (un-normalized) effects are {\em
  canonically} isomorphic to their dual cones of (un-normalized)
effects, meaning that the isomorphism defines an inner product.
Secondly, they are {\em homogeneous}: their groups of
order-isomorphisms act transitively on the interiors of their positive
cones. These two properties come close to characterizing
finite-dimensional quantum and classical state spaces: according to a
celebrated theorem, due to Koecher \cite{Koecher} and Vinberg
\cite{Vinberg}, finite-dimensional homogeneous, self-dual cones are
precisely the cones of positive elements of formally real Jordan
algebras. Once one has gone this far, two further axioms (local
tomography, and the existence of qubits) suffice to recover QM
uniquely.  Here, we establish that in any probabilistic theory in
which universal self-steering is possible (meaning that every state is
the marginal of a bipartite state steering for that marginal)
state spaces must be homogeneous and {\em weakly} self-dual, meaning
that the cones of un-normalized states and un-normalized effects must
be isomorphic, but perhaps not not canonically so. This reduces the
gap between the generic ``self-steering" theory and quantum mechanics,
largely to that between weak and strong self-duality.

A brief outline of the rest of this paper is as follows. In Section 2,
we review, for the reader's convenience, the mathematical framework in
which we work, a variant of the generalized probability theory
proposed by Mackey \cite{Mackey} over 50 years ago, and now quite
standard in foundational work in quantum theory.  In particular, we
discuss what we mean by a composite system, and by a probabilistic
``theory".  (A more detailed treatment of some of this material, can
be found, e.g., in \cite{BBLW06}.)  In Section 3, we introduce weakly
self-dual state spaces, and establish some of their properties. In
particular, we show (as a case of a result that we establish for
\emph{general} state spaces) that, for an irreducible state space $A$,
any bipartite state on $A \otimes A$ corresponding to an
order-isomorphism between $A$ and its dual is pure. \tempout{in the
  maximal tensor product of \cite{BBLW06} (and hence also in any
  smaller composite state space that contains it).} In Section 4, we
connect this to the possibility of purifying a state, showing that
{\em weakly} self-dual homogeneous state spaces are precisely those in
which every interior state (i.e., state not on the boundary of the
state space) arises as the marginal of such an isomorphism
state. Section 5 discusses steering {\em per se}, illustrating the
idea with several examples, and establishing an order-theoretic
necessary and sufficient condition for the steering of one marginal.
suitable quotient of the other. Section 6 summarizes our main results
and suggests a number of questions for further work.  

\section{The ordered linear spaces formalism}\label{sec: OLS}

This section provides a quick summary of the formalism of abstract
state spaces used in \cite{BBLW06, BBLW07, BBLW08, BW09a, BW09b}.  In
the interest of brevity, we omit detailed motivation for the
definitions below, referring the interested reader to
\cite{BBLW06}. Suffice it to say here that any model of a
probabilistic system characterized by states and observables in the
usual way \cite{DaviesLewis, Edwards}, fits naturally into this very
general framework. Although much of what we do below can be extended
to a more general context, we restrict ourselves here to
finite-dimensional systems. Thus, we assume---generally, without
further comment---that {\em all vector spaces in what follows are
  finite dimensional.} In particular, this allows us to routinely identify a vector
space $V$ with its double dual $V^{\ast \ast}$.

An \emph{ordered linear space} (OLS) is a real vector space $V$ equipped
with a partial ordering compatible with the linear structure in the
sense that it satisfies $x \leq y \Rightarrow x + z \leq y + z$ and $x
\le y \implies tx \leq ty$ for all $x,y, z \in V$ and all non-negative
scalars $t$.  Any such ordering is determined by the the pointed
convex cone\footnote{A {\em convex cone} in a real vector space $V$ is a
  convex set $K \subseteq V$ closed under multiplication by
  non-negative scalars. If $K \cap -K = \{0\}$, the cone is said to be
  {\em pointed}.} $V_+$, called the \emph{positive cone}, of vectors $x$
with $0 \leq x$, since $x \leq y$ iff $y - x \in V_{+}$; conversely,
any pointed convex cone induces an ordering on $V$ in this
way. If $a$ and $b$ are elements of an ordered linear space $V$ with $a \leq b$, we write $[a,b]$ for the set 
of vectors $x \in V$ with $a \leq x \leq b$. It is easy to see that this set is convex.

If a pointed convex cone is also generating (i.e., spans $V$, so that
$V = V_{+} - V_{+}$) and closed, it is called
\emph{regular}. Henceforth, we mean by ``ordered linear space'' one whose
positive cone is regular.  Examples include the space ${\mathbb
  R}^{X}$ of real-valued functions on a set $X$, ordered pointwise on
$X$, and the space ${\cal L}(\H)$ of Hermitian operators on a
(finite-dimensional) Hilbert space $\H$, with the usual
operator-theoretic order, i.e., $a \geq 0$ iff $a = b^{\dagger}b$ for
some $b \in {\cal L}(\H)$.  

We say that a linear map $\phi : V \rightarrow W$ between ordered
linear spaces $V$ and $W$ is {\em positive} iff it is
order-preserving, or equivalently iff it takes the positive cone of
$V$ into that of $W$, i.e., $\phi(x) \in B_{+}$ for all $x \in V_{+}$.
In particular, a linear functional $f \in V^{\ast}$ is positive iff
$f(x) \geq 0$ for all $x \in V_{+}$.  An {\em order isomorphism}
between ordered linear spaces $V$ and $W$ is a positive linear
isomorphism $\phi : V \rightarrow W$ with positive inverse---that is,
$\phi$ is an order isomorphism iff it is bijective and satisfies
$\phi(x) \geq 0$ in $W$ iff $x \geq 0$ in $V$.  An order-isomorphism
from an OLS to itself is an \emph{order-automorphism} (or just
\emph{automorphism}).  The set $\L_{+}(V,W)$ of positive linear
mappings from an OLS $V$ to an OLS $W$ is a pointed, closed convex
cone in the space $\L(V,W)$ of all linear maps from $V$ to $W$; where
$V$ and $W$ are finite-dimensional, this cone is also generating.  In
the special case where $W = {\mathbb R}$, we write $\L_{+}(V,{\mathbb
  R})$ as $V^{\ast}_{+}$, referring to this as the {\em dual cone} to
$V$.


An {\em order unit} {\em in} an ordered linear space $V$ is a vector
$u \in V_{+}$ such that, for every $x \in V_{+}$, there is some
positive scalar $t$ with $x \leq tu$. An order unit {\em on} $V$ is an
order unit {\em in} $V^{\ast}$---equivalently (in finite dimensions),
a {\em strictly} positive functional $u \in A^{\ast}$, that is, one
with $u(\alpha) > 0$ for $\alpha > 0$. (In other words, $u$ is in the
interior of $A_+^*$.)   For example, if $A = {\mathbb
  R}^{X}$, the functional $u(f) = \sum_{x \in X} f(x)$ is an order
unit. For $A = {\cal L}(\H)$, the trace is an order unit.

A {\em face} of a cone $V_{+}$ is a sub-cone $F_{+}$ with the property
that if $x, y \in V_{+}$ with $x + y \in F_{+}$, then $x, y \in
F_{+}$.\footnote{A \emph{subcone} of $V_{+}$ is, of course, a subset
  of $V_+$ that is itself a cone.} In particular, if $F_{+}$ is a face
and $0 \leq x \leq y \in F_{+}$, then $x \in F_+$ as well. The
smallest face containing a given element $y \in V_{+}$ is denoted
$\face(y)$. When this coincides with the ray generated by $y$, we say
that $y$ is {\em extremal} in $V_+$ (note this is not the same thing
as saying $y$ is an extreme point of $V_+$---only $0$ is that).  A
{\em proper} face, i.e., one not the whole cone, is contained in the
topological boundary of the cone.  An easy exercise shows that $y$ is
an order-unit in the OLS $F := F_{+} - F_{+}$ spanned by $F_{+}$, iff
$F_{+} = \face(y)$.  Note also that the intersection of faces is a
face, and that a face of a face of $V_+$ is a face of $V_+$.

If $A$ and $B$ are ordered linear spaces, there is a natural ordering
on their direct sum, namely, $(A \oplus B)_+ = \{ x + y | x \in A_{+},
y \in B_{+}\}$. We refer to $A \oplus B$, with this ordering, as the
{\em ordered} direct sum of $A$ and $B$. An ordered linear space $V$
is {\em irreducible} iff there exists no non-trivial decomposition $V$
of $V$ as an ordered direct sum. Every OLS in finite
dimension is a direct sum of irreducible ones.  An OLS is {\em
  simplicial} iff it can be represented as an ordered direct sum of
one-dimensional subspaces.

\begin{definition} By an {\em abstract state space}, we mean a pair $(A,u_A)$ where 
$A$ is an ordered linear space and $u_{A}$ is a distinguished
  order-unit on $A$. We refer to a positive element of $A$ with
  $u_A(\alpha) = 1$ as a {\em normalized state}. The set of all
  normalized states is a compact convex subset of $A_{+}$, which we
  denote by $\Omega_A$. \footnote{The set $\Omega_A$ is a {\em base}
    for the positive cone $A_{+}$: a convex set $S$ such that every
    non-zero $\alpha \in A_{+}$ is a positive scalar multiple of a
    unique vector in $S$. In the case $S=\Omega_A)$, the vector is
    $\alpha/u_A(\alpha)$. Indeed, what we
    are calling an abstract state space is essentially the same thing
    as a ordered vector space with a distinguished cone-base, i.e.,
    what we might call a (finite-dimensional) cone-base
    space.}\end{definition}

In quantum mechanics, the relevant example is $A = {\cal L}(\H)$, as
described above, with $u_A(a) = \tr(a)$; thus, $\Omega_A$ is the set
of density matrices of $\H$.  However, any finite-dimensional compact
convex set can be represented, in an essentially canonical way, as
$\Omega_A$ for a suitable abstract state space $(A,u_A)$
\cite{Shultz}, so the definition allows for arbitrarily general
models.

Of course, the picture thus far is incomplete: we also need some way to 
describe the results of measurements performed on a system. To this 
end, note that if $a$ is an outcome of some measurement, then, for any 
state $\alpha \in \Omega_A$, there should be a well-defined probability 
$\alpha(a)$ to obtain $a$ as the result of measuring the system when the 
latter is in state $\alpha$.  In order to maintain consistency with our 
intuitive notion that convex combinations of states reflect randomized 
preparations, we must require that $\alpha \mapsto \alpha(a)$ be affine, 
i.e., that it preserve convex combinations. It can be shown that any 
affine functional on $\Omega_A$ extends uniquely to a positive linear functional 
on $A$; thus, we are led to the following 

\begin{definition} An {\em effect} on an abstract state space 
$(A,u_A)$ is a positive functional $a \in A^{\ast}$ such that 
$a \leq u_A$---equivalently, $a \in A^{\ast}$ is an effect iff 
$0 \leq a(\alpha) \leq 1$ for 
every normalized state $\alpha \in \Omega_A$. \end{definition}

Note that $0$ and $u_A$ are, respectively, the smallest and largest
effects on $A$, and the set of all effects on $A$ is precisely the
interval $[0,u_A]$. From the discussion above, we see that every
measurement outcome will correspond to (or define) an effect on
$A$. We make the further assumpition here that the converse holds,
i.e, that every effect represents a measurement outcome.  \tempout{We
  understand effects as modelling measurement outcomes: if $a$ is an
  effect on $A$, we understand $a(\alpha)$ as giving the {\em
    probability} that (the outcome corresponding to) $a$ occurs in
  state $\alpha \in \Omega_A$.}  Accordingly, a discrete {\em
  observable} on $A$ is a family $\{a_x\}_{x \in X}$ of effects,
indexed by a finite set $X$ (a ``value space"), with $\sum_{x \in X}
a_x(\alpha) = 1$ for all $\alpha \in \Omega$, i.e., with $\sum_{x} a_x
= u_A$.  In the classical case where $A = {\mathbb R}^{S}$ for a
finite set $S$, an observable in this sense corresponds to a ``fuzzy"
random variable, while in the quantum case, with $A = {\cal L}(\H)$,
an effect is a positive operator between $0$ and $\1$, and an
observable is a discrete POVM (positive operator-valued measure).  A
common type of observable has $X = \{1,2,...,n\}$; such an observable
amounts to a set $a_1,...,a_n$ of effects with $\sum_{i} a_i = u_A$.

The formalism sketched above accommodates composite systems. Let $A$
and $B$ be abstract state spaces, with order-units $u_{A} \in
A^{\ast}$, $u_{B} \in B^{\ast}$ and normalized state spaces
$\Omega_{A}$ and $\Omega_{B}$. We write $A \maxtensor B$ for the space
of bilinear forms on $A^{\ast} \times B^{\ast}$, ordered by the cone
of forms nonnegative on products $a \otimes b$ of positive elements
(i.e. $a \in A^*_+$, $b \in B^*_+$), with order unit $u_{A} \otimes
u_{B}$. We write $A \mintensor B$ for the same space, ordered by the
(generally, much smaller) cone generated by the product
states $\alpha \otimes \beta$, where $\alpha \in A_{+}$ and $\beta \in
B_{+}$. 

States in $A \maxtensor B$ satisfy a natural no-signaling condition,
namely, that the marginal states of $A$ and $B$ are well-defined, not
depending on which observable may be measured on the other
wing. Conversely, it can be shown \cite{KRF, Wilce92} that a joint
probability assignment to measurement outcomes associated with $A$ and
$B$ that satisfies this non-signaling requirement, necessarily extends
to a positive bilinear form on $A^{\ast} \times B^{\ast}$, hence, to
an element of $A \maxtensor B$. Thus, the maximal tensor product
captures all non-signaling states---at least, insofar as we regard
bipartite states as determined by joint probability assignments to
pairs of local measurement outcomes.  This last assumption, sometimes
called {\em local tomography} or \emph{local observability}, is
well-known to be violated by real quantum mechanics in which the
composite of two systems described by $n$-dimensional real Hilbert
spaces is taken to be the state space of the tensor product of the two
Hilbert spaces.  Attempts to similarly describe the composite of two
quaternionic quantum systems are even more problematic: the state
space over a tensor product of quaternionic ``Hilbert spaces'' is too
small to accomodate even the product states.  Local tomography is
therefore often suggested \cite{Araki, Hardy, Barrett, BBLW06,
  D'Ariano09, Dakic-Brukner} as a possible axiom for quantum theory.
We shall adopt it here as a working assumption. 

More generally, we can consider the space $(A^{\ast} \otimes
B^{\ast})^{\ast}$ of bilinear forms on $A^{\ast} \times B^{\ast}$,
equipped with any cone lying between the minimal and maximal ones, as
``a tensor product" of $A$ and $B$. (Note that as we are in finite
dimensions, $(A^{\ast} \otimes B^{\ast})^{\ast}$ and $A \otimes B$ are isomorphic
as vector spaces.  In what follows, we shall write $AB$, generically,
for such a composite.
\footnote{We might, abandoning local tomography, 
also consider still larger spaces, of which $A
  \otimes B$ is only a quotient.  Indeed, this is necessary to
  accommodate the states on the usual tensor product of real Hilbert
  spaces, as a composite state space.}
  
If $F_{+}$ is a face of $A_{+}$, then any choice of composite $AB$ induces a
canonical choice for a composite of $F$ and $B$, which we denote by
$FB$.  It consists of those states whose $A$-marginals lie in $F$, and
is easily seen to be a face of $AB$.  For $G$ a face of $B$, $AG$ is
defined similarly.  We also write $FG$, for the face $FB \intersect AG$.  

For the purposes of this paper, we may understand by the phrase {\em
  physical theory}, a class of abstract state spaces, closed under the
formation of such a product, so as to allow the representation of
composite systems.  In a more complete treatment of this idea, one
may take a theory to be a {\em category} of abstract state spaces,
with morphisms corresponding to the processes allowed by the theory.
For some further development of this idea, see \cite{BW09a,BW09b}. In
accordance with our standing assumption, we here consider only {\em
  finite-dimensional} theories, i.e., those consisting of
finite-dimensional state spaces.

\section{Weak self-duality}

A bipartite state on a composite system $AB$, represented by a
positive bilinear form $\omega : A^{\ast} \times B^{\ast} \rightarrow
{\mathbb R}$, can also be represented by a positive map
$\hat{\omega}: A^{\ast} \rightarrow B = B^{\ast \ast}$ defined by $\hat{\omega}(a)(b)
= \omega(a,b)$. Note that we then have $\hat{\omega}(u_A) =
\omega^{B}$, the $B$ marginal of $\omega$. Notice also that the
adjoint map $\homega^{\ast} : B^{\ast} \rightarrow A^{\ast \ast} =
A$ represents the same state, evaluated in the opposite order, i.e,
$\homega^{\ast}(b)(a) = \homega(a)(b) = \omega(a,b)$.  Hence,
$\hat{\omega}^{\ast}(u_B) = \omega^{A}$. (Conversely, any positive
linear map taking $u_A$ to a normalized state of $B$ defines a
normalized bipartite state in $A \maxtensor B$.)

\begin{definition}  {\em An abstract state space $A$ is {\em weakly self-dual} iff there exists an order-isomorphism $\eta : A^{\ast} \simeq A$.}
\end{definition}

Multiplying by a sufficiently large or small positive scalar if
necessary, one can assume in the above definition that $\eta(u_{A}) =:
\alpha_o \in \Omega_{A}$, i.e., $\eta$ defines a bipartite state. It
follows that $\eta^{-1}(\alpha_o) = u_A$, so $\eta^{-1}$ is a
bipartite effect. [HB: Not clear to me this follows.  More argument
  needed...perhaps define isomorphism effect earlier and use its
  properties?]

\begin{definition} {\em A bipartite state $\omega$ in $A \maxtensor B$ is
an {\em isomorphism state} iff $\hat{\omega}: A^{\ast} \rightarrow B$
is an order isomorphism.} \end{definition}

The existence of a composite containing isomorphism states is far from
guaranteeing the weak self-duality of $A$ or $B$; indeed, for
\emph{any} state space $B$, there is a state space $A$ whose positive
cone is isomorphic to the dual of $B$'s, hence for which $A \maxtensor
B$ contains isomorphism states.  But the existence of an isomorphism
state \emph{in $A \otimes A$} obviously \emph{does} imply that $A$ is
weakly self-dual.  We call such a state an \emph{automorphism state}.

If $\omega$ is a state on $AB$, and $\tau : A \rightarrow A$ and $\eta :
B \rightarrow B$ are automorphisms of $A$ and $B$, respectively, then
$\eta \circ \hat{\omega} \circ \tau^{\ast}$ defines another
isomorphism state, with
\[(\eta \circ \homega \circ \tau^{\ast})(a)[b] = \omega(\tau a, \eta^{\ast} b).\]

\begin{theorem} \label{theorem: automorphisms are extremal}
Let $A$ be an irreducible ordered linear space. Then automorphisms of $A$ lie on extremal rays of
the cone $\L_{+}(A,A)$ of positive maps from $A$ to $A$. \end{theorem}


\tempout{ \hbcomment{It might not be hard to extend this to
    steering-for-both-marginals states, if we get a nice
    characterization of them from our characterization of
    steering-for-one-marginal states.  See further discussion after
    the results on the later.}  } 

{\em Proof of Theorem:}
Let $\chi$ be an automorphism on $A_{+}$. 
Suppose $\chi = \psi + \mu$, where $\psi, \mu : A
\rightarrow A$ are positive maps. Let $x \neq 0$ be extremal in
$A_{+}$; then $\chi(x), \psi(x), \mu(x)$ are also extremal, whence, as
$\chi(x) = \psi(x) + \mu(x)$, there are constants $c_x, d_x \geq 0$
with
\begin{equation}
\label{psi} \psi(x) = c_x \chi(x)
\end{equation}
\begin{equation}
\label{mu} \mu(x) = d_x \chi(x)
\end{equation}
\begin{equation}
\label{sum} c_x + d_x = 1
\end{equation}
We will show that $c_x$ and $d_x$ are independent of $x$, so
that $\psi = c\chi$ and $\mu = d\chi$.

It will be sufficient to prove this for $x$ ranging over the elements
of a basis $E = \{x_i\}$ for $A$ consisting of extremal elements of
$A_+$. Let $S \subseteq E$ be maximal with respect to the property
that
\[c_x = c_x' \ \text{and} \ d_x = d_x'\]
for all $x, x' \in S$. We claim that $S = E$. To see this, suppose $y$
is any extremal element of $A_{+}$, and let $S_y$ be the support of
$y$ in $E$. Expanding $y$ in the basis $E$, we have $y =
\sum_{i} t_i \chi(x_i)$.  From (\ref{psi}) and (\ref{mu}), we have
$\psi(y) = c_y \chi(y) = c_y \sum_{i} t_i \chi(x_i)$ and $\mu(y) = d_y
\chi(y) = d_y \sum_{i} t_i \chi(x_i)$, with $c_y + d_y = 1$ from
(\ref{sum}). Alternatively, expanding $y$ before applying $\psi$ and
$\mu$ gives $\psi(y) = \sum_{i} t_i \psi(x_i) = \sum_{i} t_i c_{x_i}
\chi(x_i)$, $\mu(y) = \sum_{i} t_i \mu(x_i) = \sum_{i} \alpha_i
d_{x_i} \chi(x_i)$.  Since $\{x_i\}$ is a basis and $\chi$ an
automorphism, $\{\chi(x_i)\}$ is a basis as well.  But the
expansion of an element in a basis is unique.  So $c_y \sum_{i} t_i
\chi(x_i) = \sum_{i} t_i c_{x_i} \chi(x_i)$, $d_y \sum_{i} t_i
\chi(x_i) = \sum_{i} t_i d_{x_i} \chi(x_i)$, and thus for all $i$,
either $t_i = 0$ or $c_y = c_{x_i}$ and $d_y = d_{x_i}$. Thus, if $S_y
\cap S \not = \emptyset$, the set $S_y \cup S$ again enjoys the
property that the coefficients $c_{x_i}$ and $d_{x_i}$ are constant;
since $S$ is maximal with respect to this property, $S_y \subseteq S$.
Letting $K(S)$ denote the cone $A_+ \cap \spn(S)$, we now see that
every extremal point of $A_+$ lies either in $K(S)$ or in $K(E
\setminus S)$.  That is,  $A_+$ is the direct convex sum of
$K(S)$ and $K(E \setminus S)$. As $A_+$ is irreducible, we must have
$K(E \setminus S) = \{0\}$, i.e., $S = E$. This completes the proof
\qed

\tempout {

Now, let $S_y \subset \{x_i\}$ be those $x_i$ such that for all $x_i
\in S_y$, $c_{x_i} = c_y$ and $d_{x_i} = d_y$ (and thus the $c_{x_i}$
and $d_{x_i}$ are equal for all $x_i \in S_y$). Let $z$ be an extremal
not on any ray $c x_i$ or $c y$, such that when $z$ is decomposed into
the $\{x_i\}$ basis as $z = \sum_{i} \alpha_i x_i$, there is some $x_a
\in S$ and $x_b \notin S$ such that $c_{x_a} \neq 0$, $c_{x_b} \neq
0$, $d_{x_a} \neq 0$, and $d_{x_b} \neq 0$. (If such a $z$ did not
exist, the cone would be reducible.) As shown above, $c_{x_a} = c_z =
c_{x_b}$ and $d_{x_a} = d_z = d_{x_b}$, so $T = S \cup \{x_b\}$ is
another set such that for all $x_i \in T$, $c_{x_i} = c_T$ and
$d_{x_i} = d_T$ for some $c_T$ and $d_T$. In particular, $T$ is larger
than $S$. Since we are dealing with a finite-dimensional space, we can
continue this inductively until we show that for all the $x_i$,
$c_{x_i} = c$ and $d_{x_i} = d$. This implies that $\psi(x) = c
\chi(x)$ and $\mu(x) = d \chi(x)$ for all $x$. So $\phi$ and $\mu$ are
multiples of $\chi$.  Therefore, $\chi$ is extremal.
\qed 
}

\begin{example}{\bf Automorphisms need not be extremal in reducible cones.} 
{\rm Consider the cone in two dimensions with extreme rays along the
  positive $x$ and $y$ axes. Consider the convex base with extreme
  points $(0,1)$ and $(1,0)$, and let $\chi$ be the automorphism such
  that $\chi(x,y) = (2x,y)$. Let $\phi$, $\mu$ be automorphisms such
  that $\phi(x,y) = (x/2,y/2)$, $\mu(x,y) = (3x/2,y/2)$. Then for all
  $(x,y)$, $\chi(x,y) = \phi(x,y) + \mu(x,y)$. But $\phi$ and $\mu$
  are not multiples of $\chi$. Thus the automorphism $\chi$ is not
  extremal.} 
\hfill $\bigtriangleup$
\end{example}

Note that if $A \simeq B$, say by an isomorphism $\eta : B \simeq A$,
then there is an order-isomorphism $\L(A,B) \simeq \L(A,A)$ given by
$\phi \mapsto \eta \circ \phi$, where $\phi : A \rightarrow B$. Thus,
Theorem 3.3 tells us that order-isomorphisms (if any exist) are
extremal in the cone of positive linear maps between any two
finite-dimensional ordered linear spaces. In particular, if $A$ and
$B$ are abstract state spaces, then if we interpret positive linear maps
$A^{\ast} \rightarrow B$ as bipartite states between $A$ and $B$, we
have

\begin{corollary}\label{cor: isomorphism states pure} 
If $A$ is an irreducible abstract state space, then isomorphism states
(if any exist) are pure in $A \maxtensor B$. \end{corollary}

\tempout{ \hbcomment{It might not be hard to extend this to
    steering-for-both-marginals states, if we get a nice
    characterization of them from our characterization of
    steering-for-one-marginal states.  See further discussion after
    the results on the later.}  } 

We will say a positive map $\phi: A \rightarrow B$ between ordered linear spaces $A$ and $B$ 
\emph{factors  isomorphically through} a face $F_{+}$ of $A_{+}$ if there exists a positive, 
idempotent linear map $p : A \rightarrow F$ such that $p(A_+) = F_{+}$, and $\phi = \phi' \circ p$, where $\phi' : F_+ \rightarrow Y$ is an order-isomorphism from 
$F$ onto the span of a face of $Y$.  We have the
following extension of Theorem \ref{theorem: automorphisms are
  extremal}:

\begin{corollary}\label{cor: projection isomorphism pure}
Let $\omega$ be a state in $A \maxtensor B$.  If $\homega: A^*
\rightarrow B$ factors isomorphically through an irreducible face of $A^*$ 
then it lies on an extremal ray in the cone of positive maps from $A^*$
to $B$.
\end{corollary}

\begin{proof}
Let $A$ and $B$ be finite-dimensional ordered linear
spaces with regular cones $A_+$ and $B_+$.  Suppose a positive
surjection $\phi : A \rightarrow B$ factors as $\phi = \phi' \circ p$
where $p : A \rightarrow F$ is a positive idempotent projecting $A$
onto the span of a face $F_{+}$ of $A_{+}$. Assume $\phi'$ is an
order-isomorphism, hence, extremal in $\L_{+}(F,B)$. We'd like to show
that $\phi$ is extremal in $\L_{+}(A,B)$.

To this end, let $\phi = \alpha + \beta$ where $\alpha, \beta \in
\L_{+}(A,B)$. It will be enough to show $\alpha$ and $\beta$ are
multiples of one another.  Let $x \in A$ be extremal.  We can decompose $x$ as $x = x_0 + x_1$ where $x_1 \in
\im(P)$ and $x_0 \in \ker(P) = \ker(\phi)$. Now
\[\alpha(x_0) + \beta(x_0)=  \phi(x_0) = 0.\]
Hence, $\alpha = -\beta$ on $\ker(P)$. Also,
\[\alpha(x_1) + \beta(x_1) = \phi'(x_1)\]
for all $x_1 \in \im_{+}(p)$, so, by the extremality of $\phi'$ and $x_1$, we
have
\[\alpha(x_1) = c \phi'(x_1) \ \text{and} \ \beta(x_1) = d\phi'(x_1)\]
for all $x_1 \in \im(P)$, with $c, d \geq 0$ and $c + d = 1$.  We wish
to show, then, that $\alpha(x_0) = \beta(x_0) = 0$ for all $x \in
A$---equivalently, for all $x_0 \in \ker(\phi) = \ker(p)$.
To see this, let $y \in F_{+}$ be extremal. Since $p$ takes $A_{+}$
onto $F_{+}$, we can find some $x_1 \in A_{+}$ with $p(x_1) =
y$. Since $\phi' : F_{+} \simeq B$ is an order-isomorphism, we have
$\phi(x) = \phi'(p(x_1))$ extremal in $B_{+}$ for any $x = x_1 + x_0$
with $x_0$ any element of $\ker p$.  From the discussion above, we
have
\[\alpha(x_1) = \phi(x) - \alpha(x_0) = \phi(x) + \beta(x_0), \ \text{and also} \ \beta(x_1) = \phi(x) - \beta(x_0).\]
As $\alpha$ and $\beta$ are positive maps, $\alpha(x_1)$ and
$\beta(x_1)$ lie in $B_{+}$. Since $\phi(x)$ is extremal and $\phi(x)
\pm \beta(x_0) \geq 0$, it must be that $\beta(x_0)$ is a non-negative
multiple of $\phi(x)$. A similar argument shows that $\alpha(x_0)$ is
a non-negative multiple of $\phi(x)$. But then, as $\alpha(x_0) =
-\beta(x_0)$, we must have $\alpha(x_0) = \beta(x_0) = 0$. \qed
\end{proof}

\section{Purification}

An important fact about quantum states is that they can be {\em
  purified}: any state is the marginal of a pure bipartite state. One
would like to know to what extent this is true more generally.  Within 
the general framework developed above, we can already obtain, fairly easily,  
some remarkably strong results in this direction. 

Given an abstract state space $(A,u)$, we can turn $A^{\ast}$ into an
abstract state space by using any interior {\em state} $\alpha_o \in
A_{+}$ as the order unit. We shall write $A^{\blacklozenge}$, generically, for
  such a state space $(A^{\ast},\alpha_o)$, leaving the choice of
  $\alpha_o$ tacit. Since the latter choice is, in general, not at
  all canonical, there are generally many non-isomorphic state spaces
  related to $A$ in this way, none of which has any special status as
  ``the dual" of $A$.  Nevertheless, these spaces are useful. For one
  thing, the identity map $A \rightarrow A$ can be interpreted as an
  isomorphism (hence, by Theorem 3.3, pure if $A$ is irreducible) 
state in $A^{\blacklozenge}
  \maxtensor A$ having $\alpha_o$ as its $A$-marginal. In this sense,
  every state interior to $A$ has a purification. In general, however,
  the ``ancilla" $A^{\blacklozenge}$ in terms of which $\alpha_o \in A$ is
  purified, depends on $\alpha_o$.

\tempout{Nevertheless, these spaces are useful. For one thing, an
  ensemble for $\alpha_o$ can be regarded as an observable in
  $A^{\blacklozenge}$, and the identity map $A \rightarrow A$ can be regarded as
  a bipartite state $(A^{\blacklozenge})^{\ast} \rightarrow A$ with marginal
  $\alpha_o$, which is steering for $\alpha_o$. As a rule, however,
  there will {\em not} exist bipartite states in $A^{\blacklozenge} \maxtensor A$
  steering to {\em other} states in $A$. This is one consequence of
  the following [OK -- this later...]}

\begin{theorem} \label{theorem: isomorphism homogeneity} 
The following are equivalent:
\begin{itemize}
\item[(a)] $A$ is homogeneous;
\item[(b)] Every normalized state in the interior of $A_{+}$ is the
$A$-marginal of an isomorphism state in $B \maxtensor A$, where $B$ is any 
(fixed) state space order-isomorphic to $A^{\ast}$. 
\end{itemize}
\end{theorem}

This gives us a physical interpretation of homogeneity: first, that
the various ``dual" abstract state spaces $A^{\blacklozenge} =
(A^{\ast},\alpha_o)$ are all isomorphic, not only as ordered linear
spaces but \emph{as abstract state spaces} (although not necessarily
in a canonical way), and second, as telling us that in the irreducible
case, all interior states of $A$ can be purified to isomorphism states
using a fixed ancilla, namely, any choice of
$A^{\blacklozenge}$. (Note, too, that because the dual of a homogeneous
cone is homogeneous, condition (b) is also equivalent to
the homogeneity of $A^{\ast}$.)

{\em Proof:} (a) $\Rightarrow$ (b) Consider an order-isomorphism
${\eta} : B^{\ast} \rightarrow A$.  Define ${\eta}(u_{B})
=: \alpha_o$, then since $u_B$ is in the interior of $B^{\ast}_{+}$,
$\alpha_o$ belongs to the interior of $A_{+}$.  Since the latter is
homogeneous, for any normalized state $\alpha$ we can find some
order-isomorphism $\tau : A \simeq A$ with $\alpha = \tau (\alpha_o)$;
thus, $\alpha = (\tau \circ {\eta})(u_B)$. Note that as $\alpha \in
\Omega_{A}$, it follows that $\tau \circ {\eta}$ defines a normalized
bipartite state, with marginal $\alpha$.

(b) $\Rightarrow$ (a):  Let $\alpha, \beta$ be any two
elements of $\interior A_+$.  Let $t \alpha, s
\beta$ be the normalized versions of $\alpha$ and $\beta$, with $t, s > 0$. 
Let $\omega_\alpha, \omega_\beta$ be the isomorphism states
on $B \maxtensor A$ with $A$-marginals $t \alpha, s \beta$ respectively, whose
existence is guaranteed by $(b)$.  That is, $\homega_\alpha(u_B) = t
\alpha$, $\homega_\beta(u_B) = s \beta$.  The automorphism $(s/t)
\homega_\beta \circ \homega_\alpha^{-1}$ takes $\alpha$ to $\beta$, so
$A$ is homogeneous. \qed

In the case that $A$ is weakly self-dual, we can use an
order-isomorphism $\eta : A^{\ast} \simeq A$ to identify $A$ with
$A^{\blacklozenge}$, using $\alpha_o = \eta(u_A)$. Applying the
preceding Theorem, we have

\begin{corollary} \label{cor: isomorphism homogeneity wsd} For any 
irreducible state space $A$, 
the following are equivalent:
\begin{itemize}
\item[(a)] $A$ is weakly self-dual and homogeneous;
\item[(b)] Every normalized state in the interior of $A_{+}$ is the
  marginal of an isomorphism state in $A \maxtensor A$.
\end{itemize}
\end{corollary}

\noindent{\em Remark:} 	Recently, Chiribella, D'Ariano and Perinotti
\cite{D'Ariano09} have examined the consequences of assuming, as an
axiom, that all states dilate to a pure state that is unique up to a
reversible transformation on one marginal.  We note that if $\alpha
\in \Omega_A$ can be achieved as the marginal of two isomorphism
states $\omega$ and $\mu$ in the same composite $BA$ (so $B^* \iso A$,
say via an isomorphism $\sigma$), so that $\alpha = \homega(u_B) =
\hat{\mu}(u_B)$, then $\tau := \homega \circ \hat{\mu}^{-1}$ is a
unit-preserving order-automorphism of $B^*$, and $\omega(a,b) =
\mu(\tau(a),b)$.  We are using the convention $\homega: B^*
\rightarrow A$; $\tau$ is a reversible map acting on $B^*$ (note that
unit preservation is the condition dual to base preservation).  So for
the set of isomorphism states having $\alpha$ as $A$-marginal, the
dilation condition is met; in constructing a composite, uniqueness can
be ensured by including no other pure states with $\alpha$ as
marginal, in the extreme generators of the composite cone.


\tempout {\begin{problem} Does there exist a homogeneous, weakly
    self-dual, but not self-dual state space?  If so, exhibit one.
\end{problem}}

\section{Steering}\label{sec: steering}

By an {\em ensemble} for a state $\beta \in B$, we mean a finite set
of $\beta_i \in B_+$ such that $\sum_i \beta_i = \beta$.  Note that we
defined ensembles not as lists of probabilities and associated
normalized states, but as lists of unnormalized states; the two
definitions are equivalent, as the norms
\footnote{Although we will not need it, there is a unique norm, called
  the base norm, that agrees with the linear functional $u_B$ on $B_+$,
  so the terminology is justified.} $u_B(\beta_i)$
of the $\beta_i$ encode the probabilities.  Indeed, from $\sum_i
\beta_i = \omega^B$ and the positivity of the $\beta_i$, it follows
that $u_B(\beta_i)$ must be probability weights, $0 \le u_B(\beta_i)$,
$\sum_i u_B(\beta_i) = 1$.  If instead $\sum_i \beta_i \le \omega^B$ is 
required, we have a \emph{subensemble} for $\omega^B$.  

As discussed in Section \ref{sec: intro}, pure quantum-mechanical states have the
interesting property that any ensemble for either marginal state can
be realized as the conditional states arising from a suitable choice
of observable on the other wing of the system.  Generalizing, we are
led to the following

\begin{definition}{\em  A bipartite state $\omega \in A \maxtensor B$ 
is {\em steering for its $B$ marginal} iff, for every ensemble (convex
decomposition) $\omega^B = \sum_i \beta_i$, where $\beta_i$ are
un-normalized states of $B$, there exists an observable $E = \{x_i\}$
on $A$ with $\beta_i = \hat{\omega}(x_i)$. We say that $\omega$ is
{\em bisteering} iff it's steering for both marginals.}
\end{definition}

If $\alpha$ is any state on $A$ and $\beta$ is a {\em pure} state on
$B$, then $\omega = \alpha \otimes \beta$ is trivially steering for
$\omega^{B} = \beta$ since the latter has no non-trivial
ensembles.  If $\alpha$ is mixed, then $\omega$ will {\em not} be
steering for $\omega^A$. On the other hand, any pure product state is
steering, as is any isomorphism state. In light of Theorem 
\ref{cor: isomorphism homogeneity wsd}, this
might suggest that steering states are always pure.  But that is not
correct. Indeed, any classical bipartite state exhibiting perfect
correlation between its marginals is steering for both its marginals.
(This is closely related to Example 2.4 above.) 
\tempout{
However, as we shall
presently show, where both marginals lie in the interiors of their
respective cones, and where the latter are irreducible, steering
states are indeed pure.
}

\tempout{ \hbcomment{Where the support of each marginal has the same
    cardinality, perfect correlation must be the precise condition for
    steering both marginals of a classical state, and we should show
    this someplace.  Finally, we should specialize the main theorem to
    the classical case, following the theorem someplace, as an example
    and aid to understanding.  AW: Agreed, but maybe not in this
    draft.}  } 

\tempout{ Consider the set of states on $AB$ steering for their $B$
  marginal.  It's obvious this is not convex: for example, an equal
  mixture of a perfectly correlated and a perfectly anticorrelated
  state of two classical bits is the maximally mixed state of two
  bits, which is a product of mixed states, and useless for steeering.
  Another example is the fact that the convex hull of the four Bell
  states is the maximally mixed state.}

It follows almost immediately from the definition, that if $\omega$ is
steering for its $B$-marginal, $\homega(A_{+})$ is a face of
$B_{+}$. Indeed, we have

\begin{lemma} \label{lemma: steering face}
If $\omega$ is steering, then $\homega(A_+) = \face(\omega^B)$. \end{lemma} 

{\em Proof:} Suppose that $\beta_1, \beta_2 \in B_{+}$ with $\beta_1 +
\beta_2 \in \homega(A_{+})$---say, $\beta_i = \homega(a_i)$ where
$a_i \in {A^{\ast}}_{+}$. Since $u_A$ is an order unit for
$A^{\ast}_{+}$, we can find a scalar $t > 0$ such that $t (a_1 + a_2)
\leq u_A$, whence,
\[t\beta_1 + t\beta_2 = \homega(t (a_1 + a_2)) \leq \homega(u_A) = \omega^B.\]
Thus, $\{t\beta_1, t\beta_2\}$ is a sub-ensemble for $\omega^B$. Since
$\omega$ steers $\omega^B$, $t\beta_1$ and $t\beta_2$---and hence,
also $\beta_1$ and $\beta_2$---lie in $\homega(A_+)$.  This shows
that $\homega(A_+)$ is a face of $B_+$. Notice that the argument also
shows that $\omega^B$ is an order unit for $\homega(A_+)$; hence, the
latter is exactly $\face(\omega^B)$. \qed

The converse, however, is not true: a state may satisfy $\homega(A_+)
= \face(\omega^B)$ but not be steering, as in the following example.   

\tempout{The following example shows that, for $\omega$ to steer for
$\omega^B$, it is {\em not} sufficient that $\hat{\omega}$ map
$A^{\ast}_{+}$ onto the face of $B_{+}$ generated by $\omega^B$; the
order-unit isomorphism condition $\homega ([0,u_A]) = [0,\omega^B]$
may still fail in this case, even though (as always) $\homega(u_A) =
\omega^B$.} 

\begin{example}
{\rm $A_+$ is the simplicial cone $\R^3_+$, $B_+$ is $\R^2_+$, with
  the usual order units $(1,1,1)$ and $(1,1)$ respectively, and
  $\omega(a,b)$ is determined by the following table of probabilities:
  \beq
\begin{array}{ccc}  
 & x & y \\ x & 1/4 & 0 \\ y & 0 & 1/4 \\ z & 1/4 & 1/4
\end{array}
\eeq The row index ranges over the values $x := (1,0), y := (0,1)$ for
$a$, the column index $b$ ranges over the values $x := (1,0,0)$, etc..

We see that $\homega(x) = (1/4, 0)$, $\homega(y) = (0, 1/4)$, and
$\homega(z) = (1/4, 1/4)$, while $\homega(u) = \homega(x + y + z) =
(1/2, 1/2)$.  The image of the order interval $[(0,0,0), (1,1,1)]$ of
$\R^3_+$ under $\homega$ is the hexagon with vertices $(0,0)$, $(1/4,
0)$, $(0, 1/4)$, $(1/2, 1/4)$, $(1/4, 1/2)$, $(1/2, 1/2)$.  This is a
proper subset of the interval $[0, \pi((1,1,1))]$, which is the square
with vertices $(0,0)$, $(1/2, 0)$, $(0, 1/2)$, $(1/2, 1/2)$.  So as
claimed, even though $\pi(u)$ is the desired order unit in $\R^2_+$,
the image $\pi([0,u])$ of the unit interval in $\R^3$, while it
generates $\R^2_+$, is not a unit interval (for {\em any} ordering of
${\mathbb R}^{2}$). 

In particular, the ensemble $(0,1/2)$, $(1/2,0)$ for $\omega^B =
(1/2,1/2)$ can not be represented as $\homega(a_1), \homega(a_2)$ for
any elements $a_1, a_2 \in [0,u_A]$---much less with $a_1, a_2$
summing to $u_A$. Accordingly, $\omega$ is not steering.
  }
\hfill $\bigtriangleup$
\end{example}

In fact, the condition that $\omega$ be steering for its $B$-marginal
places a very strong and subtle constraint on $\homega$. 
If $X$ and $Y$ are partially ordered sets, an order-preserving surjection 
$p : X \rightarrow Y$ is a {\em quotient map} iff, for all $y_1, y_2 \in X$, 
$y_1 \leq y_2$ iff $y_i = p(x_i)$ for some $x_1 \leq x_2$ in  $X$. We 
shall say that $p$ is a {\em strong quotient map} iff it has the 
property that every chain $y_1 \leq y_2 \leq \cdots \leq y_n$ in $Y$ 
is the image of some chain $x_1 \leq x_2 \leq \cdots \leq x_n$ in $X$, 
i.e., $y_1 = p(x_1), y_2 = p(x_2), ..., y_n = p(x_n)$. Evidently, 
a strong quotient map is a quotient map (just apply the definition to
chains of length $2$), but the converse is, in general, false. 

\begin{theorem} \label{theorem: bipartite quotient}
Let $\omega$ be a bipartite state in $AB$. Then $\omega$ is steering
for its $B$ marginal iff $\homega : [0,u_A] \rightarrow [0,\omega^B]$
is a strong quotient map of ordered sets.\end{theorem}

{\em Proof:} First, suppose $\omega$ is steering, and let $0 \leq
\beta_1 \leq \beta_2 \leq \cdots \beta_{k-1} \omega^B$.  Then
$\{\beta_1, \beta_2-\beta_1, \dots , \beta_{k-1}- \beta_{k-2},
\omega^B - \beta_{k-1}\}$ is an ensemble for $\omega^B$; as $\omega$
is steering, there is an observable $\{a_1,a_2,a_3,...,a_k\}$ on $A$
with $\homega(a_1) = \beta_1, \homega(a_2) = \beta_2 - \beta_1, \dots
, \homega(a_k) = \omega^B - \beta_{k-1}$. In particular, defining $b_j
= \sum_{i \le j} a_j$, we have $\homega(b_j) = \beta_j$; since 
$b_1 \le b_2 \le \cdots \beta_k$, $\homega$ induces a quotient map of
ordered sets $[0,u^A] \rightarrow [0,\omega^B]$.  For the converse,
suppose $\homega$ induces such a quotient map.  For any ensemble
$\{\eta_1,...,\eta_k\}$ for $\omega^B$, define $\beta_1 = \eta_1$,
$\beta_2 = \eta_2 + \eta_1$, ..., $\beta_j = \sum_{i \leq j} \beta_i$
for $j \leq k-1$.  By definition of a strong quotient map, there are
then elements $b_1 < b_2 < ... < b_{k-1}$ in $[0,u^A]$ with
$\homega(b_i) = \beta_i$.  Setting $a_j = b_j - (\sum_{i < j} b_i)$,
we have $\homega(a_j) = \eta_j$ and $\sum_{j \leq k-1} a_j =
b_k$. Thus, $\{a_1,...,a_{k-1},a_k := u_{A}-b_k\}$ is the desired
observable steering to $\{\eta_i\}$.  \qed

{\em Remarks:} (i) We suspect, but so far have been unable to prove, that 
a quotient map of order-intervals $[0,u] \rightarrow [0,v]$ is necessarily 
a strong quotient. 

(ii) An obvious {\em sufficient} condition for $\homega :
[0,u_A] \rightarrow [0,\omega^B]$ to be a quotient map of ordered sets is for there to
exist an affine section $\sigma : [0,\omega^B] \rightarrow
[0,u^A]$. However, as Example \ref{ex: positive section not necessary} 
in Appendix A shows, this is not
necessary for steering.

It follows from Theorem \ref{theorem: bipartite quotient} that the
ordering of $\face(\omega^B) = \homega(A_{+})$ is exactly the quotient
linear ordering induced by the linear surjection $\homega$, i.e.,
$\beta_1 \leq \beta_2$ in $\face(\omega^B)$ iff $\beta_i =
\homega(a_i)$ for some $a_1, a_2 \in A$ with $a_1 \leq a_2$.  It also
follows that if $\homega$ is injective, it is an order
isomorphism. This last point is important enough to record as

\begin{corollary} \label{cor: injective steering pure}  et $\omega$ be steering for $\omega^B$, 
where $\omega^B$ is interior to $B_{+}$, so that $\face(\omega^B) =
B_+$. If $\homega$ is injective (non-singular), then $\homega$ is an
order isomorphism.  If $B_+$ (and therefore $A_+$) is irreducible,
therefore, by Theorem \ref{theorem: automorphisms are extremal}, it is
pure in $A \maxtensor B$. \end{corollary}

In other words, if $A$ and $B$ have the same dimension, then the
states that are steering for an interior marginal are precisely the isomorphism
states (and hence steering for both marginals). 

We are now in a position to make good on the claim made in the
introduction.  

\begin{definition}
A probabilistic theory \emph{supports
  universal steering} if, for every system $B$ in the theory and every
state $\beta \in B$, there exists a system $A_\beta$ and a bipartite
state $\omega$ in $A_\beta \otimes B$ that steers its $B$-marginal
$\omega^B=\beta$.  A theory \emph{supports uniform universal
  steering} if, for every system $B$ in the theory, there exists a
system $A_B$ such that for every state $\beta \in A$, there exists a
state $\omega$ in $A_B \otimes B$ that steers its $B$-marginal
$\beta$.  A probabilistic theory {\em supports
  universal self-steering} if, for every system $A$ in the theory,
every state $\alpha \in A$ can be represented as a marginal of some
bipartite state on two copies of $A$---that is, some state $\omega
\in AA$---steering for that marginal. (That is, it supports uniform
universal steering with $A_B \iso A$.)
\end{definition}

Corollary \ref{cor: injective steering pure}, combined with Theorem \ref{theorem:
  isomorphism homogeneity}, establish

\begin{proposition}
In any theory that supports universal uniform steering, every
irreducible, finite-dimensional state space in the theory is
homogeneous.
\end{proposition}

In light of Corollary \ref{cor: isomorphism homogeneity wsd}, we also have

\begin{proposition}
In any theory that supports universal self-steering, every irreducible,
finite-dimensional state space in the theory is homogeneous and weakly
self-dual.
\end{proposition}

If a theory supports universal self-steering, and also
has the property that every direct summand of a state space is again a
state space belonging to the theory (a reasonable requirement, at
least in finite dimensional settings), then every finite-dimensional
state space in the theory is a direct sum of homogeneous, weakly
self-dual factors, hence, homogeneous and weakly
self-dual.\footnote{It is easily seen that direct sums of homogeneous
  or weakly self-dual cones are, respectively, homogeneous or weakly
  self-dual.  For weak self-duality, one just proves that any sum
  of isomorphisms $\alpha_i: A_i \rightarrow B_i$ of direct summands,
  is an isomorphism of the direct sums $\oplus_i A_i$ and $\oplus_i
  B_i$.  For homogeneity, one uses the fact that the interior of the positive cone of a
  direct sum consists precisely of sums of interior points of the positive cones of
all
  summands.  Then to get from any such interior point $x = \sum_i x_i$
  to any other $y = \sum_i y_i$, one uses a sum of automorphisms
  $\alpha_i$ of the summands $A_i$, chosen (as homogeneity of each
  $A^i_+$ ensures is possible) such that $\alpha_i(x_i) = y_i$.}

An interesting question is to what extent the gap between universal
steering and uniform universal steering is a genuine one.  One might
investigate this question by looking for examples of state spaces for
which each state can be steered, but that are not homogeneous.


\tempout { In Theorem 5.4, the ordered linear space $F =
  \face(\omega^B) - \face(\omega^B)$ generated by $\face(\omega^B)$
  can be regarded in two ways. On one hand, it can can be regarded as
  an abstract state space, with order-unit $u_B$ (or rather, the
  restriction of $u_B$ to $F$), and also as a {\em dual} state space,
  i.e., an order-unit space, taking $\omega^B$ as an order unit in
  $F$.  If we let $p : A^{\ast} \rightarrow F$ denote the
  co-restriction of $\homega$ to its range, $F$, and $i : F
  \rightarrow B$, the natural inclusion map, then the former is a map
  of order-unit spaces, and the latter is a map of abstract state
  spaces. We have, then, a factorization

\tempout{To distinguish between these two roles, let $C$ denote the
  abstract state space $(F^{\ast},\omega^B)$, so that, regarded as an
  order-unit space, $F = C^{\ast}$.  Notice that the identity map
  $\id_{A} : F \rightarrow F$ becomes, in this optic, an isomorphism
  state on $C \otimes F$, and theorem 5.4 tells us that a bipartite
  state $\omega$ is steering iff the map the map $\homega : A^{\ast}
  \rightarrow B$ factors into a surjective map $p : A^{\ast}
  \rightarrow C^{\ast}$, an isomorphism state $C^{\ast} \simeq F$, and
  the canonical inclusion map $i : F \rightarrow B$ (an embedding of
  state spaces).  Conversely, of course, any such factorization yields
  a steering state, provided that $[0,u_C]$ is the quotient of
  $[0,u_A]$ under $p$, as required by Theorem 5.4
\[ {\xymatrix@=12pt{ A^{\ast} \ar@{->}^{\hat{\omega}}[rr] \ar@{->}_{\pi}[dd] &  & B \\
& & & \\
 C^{\ast} \ar@{->}_{\phi}^{\simeq}[rr] & & F(\omega^{B}) \ar@{->}^{\hat{i}}[uu]}}\]
 }

\[ {\xymatrix@=12pt{ A^{\ast} \ar@{->}^{\hat{\omega}}[rr] \ar@{->}_{\pi}[ddrr] &  & B \\
& & & \\
 & & F(\omega^{B}) \ar@{->}^{\hat{i}}[uu]}}\]
(where, indeed, the identity map $\id_{F} : F \rightarrow F$ can be regarded as a bipartite isomorphism state 
on $F$.)
}


\tempout{ 
\begin{corollary} \label{cor: bisteering isomorphism} 
Let $\omega \in AB$ steer for both $\omega^A$ and $\omega^B$. 
Then $F(\omega^A)$ is canonically isomorphic, as an ordered linear
space, to $F(\omega^B)^{\ast}$. \end{corollary}

{\em Proof:} Since $\omega$ is steering for its $B$-marginal, we have a
positive surjection $\homega : A^{\ast} \rightarrow F(\omega^{B})$,
taking $A^{\ast}_{+}$ onto the face
$F_{+}(\omega^B)$. Dualizing, we have a positive linear injection
$\homega^{\ast} : F(\omega^{B})^{\ast} \rightarrow A$.\footnote{As discussed in
Section \ref{sec: OLS}, this simply represents the state $\omega$, evaluated in
reverse order (that is, first on effects in $B^{\ast}$, then on
effects in $A^{\ast}$).} Since $\omega$ is steering for its
$A$-marginal, Lemma \ref{lemma: steering face} tells us that the positive image of
$\homega^{\ast}$ is exactly the face $F_{+}(\omega^A)$ generated by
$\omega^A$. Thus, $\homega^{\ast}$ provides a positive linear
bijection $F(\omega^B)^{\ast} \rightarrow F(\omega^A)$.  To 
show that $\homega^{\ast}$ is an order-isomorphism, it remains only
to show in addition that 
if $b \in F(\omega^B)^{\ast}$ with ${\homega^{\ast}}(b) =: \alpha \in
A_{+}$, then $b \geq 0$. But if $\alpha \geq 0$, then for any $a \in
A^{\ast}_{+}$, we have
\[b(\homega(a)) = \omega(a,b) = \homega^{\ast}(b)(a) = 
a(\homega^{\ast}(b)) = a(\alpha) \geq 0.\]
Since $\homega$ takes $A_{+}$ onto $F(\omega^{B})_{+}$, we have that
$b$ is non-negative on the latter cone, whence, $b \in
F(\omega^B)^{\ast}_{+}$, as required. \qed
} 

\tempout{
\begin{corollary}\label{cor: bisteering pure}
With notation as above, if $\omega$ is bisteering and $\face(\omega^A)$
and $\face(\omega^B)$ are irreducible faces of $A$ and $B$, respectively,
then $\omega$ is pure in $AB$. \end{corollary}

{\em Proof: } Observe that if $F$ and $G$ are faces of $A$ and $B$,
respectively, then $FG = AG \cap FB$ is a face of $AB$. As
$F(\omega^A)$ and $F(\omega^B)$ are irreducible, Corollary 
\ref{cor: projection isomorphism pure} [CHECK THAT THIS REFERENCE IS CORRECT
AND THAT THE NEEDED CONDITIONS HOLD]
  implies that $\phi$ is pure in $F(\omega^A) \maxtensor
  F(\omega^B)$. But the latter is a face of $A \maxtensor B$.  A state
  pure in $F(\omega^A) \maxtensor F(\omega^B)$, such as $\phi$, is
  also pure in $A \maxtensor B$---and hence, in $AB$ as well. \qed
}

{\em Remark:} A particularly strong ``steering" axiom would require
that, for every state $\alpha$ of every system $A$ in the theory,
there exist a steering state $\omega$ on a composite $AA$ of two
copies of $A$, having {\em both} marginals equal to $\alpha$.  
\tempout{
This
has a number of interesting consequences, among them that every {\em
  face} of every system in such a theory is self-dual. 
}
Such a theory
must be ``mono-entropic" in the sense that measurement and mixing
entropies of states coincide, as discussed in \cite{BBCLSSWW09};
states in such a theory must also be spectral, in the sense of
\cite{Wilce09}. Further elaboration of these points can be found in
Appendix B of \cite{BBCLSSWW09}. 

\section{Conclusion and discussion} 

We have shown that the state spaces of any probabilistic theory that
allows for uniform universal ensemble steering, in the sense that for
every system $A$ in the theory, there's another system $B$ such that
every state on system $A$ can be steered by some state in the
composite $BA$, are homogeneous.  We say that system $B$ \emph{steers
  system} $A$, in this case.  If we require systems to be
\emph{self}-steering (i.e., that each system $A$ steer \emph{itself}),
they must be homogeneous and weakly self-dual. If one could motivate
the stronger assumption that these state spaces are {\em strongly}
self-dual, then the Koecher--Vinberg Theorem \cite{Koecher, Vinberg},
together with the Jordan--von Neumann--Wigner classification theorem
\cite{JvNW}, would imply that all state spaces are those of formally
real Jordan algebras.  In our finite dimensional setting, this means
that their normalized state spaces are convex direct sums
of sets affinely isomorphic to the unit-trace
elements in the cones of positive semidefinite matrices in a real,
complex, or quaternionic matrix algebra, or to Euclidean balls, 
or to the unit-trace $3 \times 3$ positive
semidefinite matrices over the octonions.

From here, our standing assumption of local tomography (that bipartite
states are determined by the probabilities they assign to product
effects) restricts the possibilities further.  A theorem of
Hanche-Olsen \cite{Hanche-Olsen} asserts that any JB-algebra $A$
(which includes all formally real Jordan algebras, at least in finite
dimension) whose vector-space tensor product with the self-adjoint
part of $M_2({\mathbb C})$---that is, with a qubit---can be made into
a JB tensor product, is isomorphic to the self-adjoint part of a
(complex) $C^*$-algebra.  In other words, it is essentially
quantum-mechanical.  \tempout{By saying that $A \otimes M_2(\C)$ is a
  JB-tensor product, we mean that it can be equipped with a Jordan
  product $\circ$ such that for all $x,y \in A$, $a,b \in M_2(\C)$,
  \beqa (x \otimes 1) \circ (y \otimes 1) = (x \circ y) \otimes 1
  \\ (1 \otimes a) \circ (1 \otimes b) = 1 \otimes (a \circ b) \\ (1
  \otimes a) \circ (x \otimes 1) = x \otimes a \eeqa and Jordan
  multiplication by $1 \otimes a$ commutes with Jordan multiplication
  by $x \otimes 1$.}  As we will establish elsewhere,
Hanche-Olsen's requirements for a JB tensor product impose 
on the cones associated with the three JB-algebras in question, exactly the 
operational requirements we've imposed on a composite of state spaces. 
Thus, Hanche-Olsen's result implies that if a homogeneous, self-dual 
state space has a \emph{locally tomographic} homogeneous, self-dual 
composite with a qubit, then it is the state space of a $C^*$-algebra 
--- so, a direct sum of the state spaces of standard quantum theory.

Therefore, if a \emph{self-dual} physical theory makes room for
qubits, and permits universal self-steering, its state spaces are
essentially quantum-mechanical, in the fairly standard sense 
that uses the complex field, but extended to allow for
superselection sectors (direct summands).  

These considerations trace a route, within a broad landscape of
locally tomographic non-signaling theories, from the single
information-theoretic feature of quantum states that most puzzled
Schr\"odinger---the possibility of steering---to the full mathematical
apparatus of (finite-dimensional) $C^*$-algebraic quantum mechanics.
This route is interrupted by a gap: that between weak and strong
self-duality.  There may be ways to bridge this gap.  One strategy for
doing so can be found in \cite{Wilce09}.

On the other hand, it would also be interesting to see whether a
complete and self-consistent theory can be constructed using only {\em
  weakly} self-dual state spaces, that still allows for universal
steering.  An essential step towards constructing such a theory would
be to find a class of weakly, but not strongly, self-dual, homogeneous
state spaces that is closed under some reasonable non-signaling
tensor product.  A plausible idea is to include \emph{all} steering
states, but as discussed in Appendix B, it does not work.  We are
investigating other possibilities based on keeping a rich supply of
steering states.  Of course, should it prove that {\em any} category
of homogeneous, weakly-self-dual state spaces that admits a reasonable
tensor product must be {\em strongly} self-dual then the gap mentioned
above will turn out to have been only an illusion, and quantum theory,
in the $C^*$-algebraic sense, will be naturally characterized, at
least in finite dimension, in terms of steering and the existence of a
locally tomographic nonsignaling tensor product.\footnote{We note that
our notion of tensor product is less innocuous than might be evident,
as it requires the positivity of bipartite 
states on \emph{all} product effects, which, for example, need not 
be required in a theory in which only a subcone of $A_+^*$ is considered
to represent operationally relevant effects on which positivity is to be
required.}


Finally, we mention that additional information-processing properties
provide some motivation for weak self-duality, both via steering, and
more directly.  For example, in \cite{BDLT2008} it was shown that an
exponentially secure bit commitment protocol, based on the
nonuniqueness of convex decomposition in nonclassical state spaces,
exists in any theory which has at least some nonclassical state
spaces, coupled only by the minimal tensor product (so that there is
no entanglement).  But in a nonclassical theory in which all states
can be steered, this type of bit commitment protocol cannot exist.
This does not provide a tight connection between steering and
no-bit-commitment; other types of bit commitment protocols might be
able to coexist with steering.  But it is suggestive.  Another
connection is between weak self-duality and teleportation: if it is
possible, to conclusively teleport a system through a copy of itself
with nonzero success probability, then it must be weakly self-dual
\cite{BBLW08}.\footnote{This also requires that all
  composites $AB$, $A'B'$ of isomorphic systems $A \iso A'$, $B \iso
  B'$, contained in the theory, be isomorphic.}

\section*{Acknowledgments}
{This research was supported by the United States Government through
  grant OUR-0754079 from the National Science Foundation.  It was also
  supported by Perimeter Institute for Theoretical Physics; work at
  Perimeter Institute is supported in part by the Government of Canada
  through Industry Canada and by the Province of Ontario through the
  Ministry of Research and Innovation.  HB and AW also wish to thank:
  Samson Abramsky and Bob Coecke for extending them the hospitality of
  the Oxford University Computing Laboratory during November of 2009,
  when parts of this paper were written; C. Martin Edwards for
  referring us to the work of H. Hanche-Olsen; the Foundational
  Questions Institute and and the University of Cambridge's DAMTP for
  sponsoring the workshop Operational Probabilistic Theories as Foils
  to Quantum Theory, Cambridge (U.K.), July 2007, where some of these
  ideas were initially presented and their development into the
  present paper begun.  Further work was done at other workshops and
  conferences: we thank Renato Renner, Oscar Dahlsten, and the the
  Pauli Center for Theoretical Studies and the initiative ``Quantum
  Science and Technology'' at ETH Z\"urich, for sponsoring the
  workshop ``Information Primitives and Laws of Physics'', March,
  2008; Jeffrey Bub, Robert Rynasiewicz, and the University of
  Maryland, College Park and Georgetown University for organizing and
  sponsoring New Directions in the Foundations of Physics, May 2008;
  Hans Briegel, Bob Coecke, and the other organizers and the European
  network QICS, the EPSRC, and the IQOQI of the Austrian Academy of
  Sciences for the workshop ``Foundational Structures for Quantum
  Information and Computation'', September 2008.}

\appendix

\section{Examples for Section 5}

\begin{example}
{\bf An example where there is a section that enables steering.}  {\rm
  Consider the following state in the maximal tensor product of two
  state spaces with square base.  We'll view this as the state space
  of two two-outcome tests $\{a, a'\}$ and $\{b, b'\}$, and also
  identify $a, a', b, b'$ with the atomic effects in the dual cone.
  We'll label each of the four vertices of the normalized state space
  by the two atomic effects it makes certain: $ab, ab', a'b,
  a'b'$. 
Writing, e.g. $ab \otimes a'b'$ for a bipartite
  product state, the state: \beq
\label{eq: first square base example} \omega = \half (ab \otimes
ab + a'b \otimes a'b) \eeq has\beq \label{images in the square base
  example} \homega(a) = \half ab, ~~ \homega(a') = \half a'b \;, \eeq
so measuring $\{a, a'\}$ on the first system gives the ensemble
$\{\half ab, \half a'b\}$ for the second-system marginal $\omega^B =
\half (ab + a'b)$.  $\face(\omega^B)$ is generated as nonnegative
linear combinations of $ab$ and $a'b$, and is thus a two-dimensional
ordered subspace of the three-dimensional state space of the second
system.  Its unit interval is the square that is the convex hull of
$(0,0), ab/2, a'b/2, (ab + a'b)/2$.  The quotient of the first system
space, $A$, by the kernel of the linear map $\homega$, can be
represented by setting up the state space to have $90^\degree$ opening
angle between opposite extremal rays, and taking the orthogonal
($90^\degree$) projection onto the plane normal to the ray $b'$,
i.e. the projection along the ray generated by $b'$. 
This
indeed gives a two-dimensional classical state space, i.e.  one
isomorphic to $\face(\omega^B)$.  Moreover, there is an affine section
$\sigma$ of this quotient map into $A^*_+$, given by inverting the
relations (\ref{images in the square base example}), thus allowing us
to map the unit interval $[0, \omega^B]$ in $\face(\omega^B)$ to the
diagonal cross section, in the $a, a'$ plane, of the unit interval of
$A^*_+$.  $\sigma \circ \pi$ is indeed a positive projection on
$A^*_+$, namely the orthogonal projection onto this plane.  }
\hfill $\bigtriangleup$
\end{example}

For the state (\ref{eq: first square base example}), the mentioned
section is the only affine section over $\face(\omega^B)$.  A slight
modification, however, gives an example in which many affine sections
exist.
\begin{example}
{\bf An example in which there are many affine sections over the image
  of the order unit.}  {\rm Let \beq \homega = \half ( x \otimes ab +
  y \otimes a'b) \eeq with $x$ any state in the line segment $[ab,
    ab']$ and $y$ any state in the segment $[a'b, a'b']$ respectively.
  This still gives a state steering for the same marginal $\omega^B
  = \half(ab + a'b)$ as in the preceding example.  If we choose $x = ab, y = a'b'$, that is,
  \beq\label{eq: second squares example} \homega = \half ( ab \otimes
  ab + a'b' \otimes a'b)\;, \eeq we can steer the marginal using
  either of the two observables $\{a,a'\}, \{b,b'\}$.  In this case,
  there are two distinct sections over $\face (\omega^B)$ that enable
  steering.  }
\hfill $\bigtriangleup$
\end{example}

In the example of Eq. (\ref{eq: second squares example}), the kernel
of $\homega$ is generated, not by $b'$ as before, but by $a'b - ab'$,
and the natural way of attempting to represent the quotient map,
namely by projection onto a subspace complementary to the kernel in
the original space, if we project orthogonally in the natural
geometry, projects onto the subspace spanned by $ab, a'b'$, and the
order unit, giving a slice of the order interval that includes the
diagonal of the square.  

Although the kernel of this map is of course one-dimensional, its
positive kernel, which we've claimed must be an exposed face of the
state cone, is the trivial such exposed face: the zero subspace.  This
example shows that quotients whose positive kernel is trivial are not
necessarily uninteresting or ill-behaved.

\begin{example}\label{ex: positive section not necessary}
{\bf An affine section of the order-quotient map is not necessary for
  steering.}  {\rm Let $A \iso \R^4$ be an abstract state space whose normalized
states form a cube; its dual cone is a regular polyhedral cone in $\R^4$ with
  octahedral base.  The atomic effects (largest effects on extremal
  rays of the effect cone) are the vertices of such an octahedral
  base.  The example has $B \iso \R^3$ ordered by a cone with regular hexagonal
  base; the center of the hexagon is a state having three distinct
  two-state extremal ensembles, each composed of a pair of opposite
  vertices of the hexagon (with weights $1/2$ on each one).
  $\homega^{AB}$ is chosen to take the six vertices of the octahedron
  of atomic effects in $A^*$ to the six vertices of the hexagon of
  states normalized to $1/2$, in such a way that opposite vertex-pairs
  are mapped to opposite vertex-pairs.  
This is clearly
  positive and linear, and maps the order-unit (twice the center of
  the octahedron of atomic effects) to the center of the hexagon of
  normalized states (which is twice the center of the hexagon of
  states normalized to $1/2$).  $\face(\homega^B)$ is the entire
  hexagonal cone in $\R^3$, so $\homega$ is itself (technically, is a
  representative of) the quotient $\pi$.  All three ensembles for
  $\omega^B$ consist of extremal states, and because they have unique
  $\homega$-preimages, a section of $\homega$, we know that a section
  of $\pi$ must take these, the vertices of a hexagon, to the vertices
  of the octahedron, matching opposites to opposites.  But no affine
  map can do this.  The affine span of the hexagonal points has affine
  dimension $2$, while that of the octagonal ones has affine dimension
  $3$.}
\hfill $\bigtriangleup$
\end{example}

\section{The Steering Product} 

A physical {\em theory}, as distinct from a model of a single physical
system, should allow some device whereby several such models can be
combined to yield a model of a composite system. It is natural to
suppose that such a theory constitutes a category equipped with a
well-behaved tensor product; that is, a symmetric monoidal category
as described in e.g. \cite{Abramsky-Coecke}. This suggests the following
problem: given two weakly self-dual state spaces $A$ and $B$, does
there exist a reasonable model for a composite state space $AB$ that
is again weakly self-dual?
In light of the connection between weak self-duality
and steering, one might consider building a weakly self-dual tensor 
product by including all steering states, but as we'll see, this
does not work, even where the factors are quantum-mechanical.

Note that for non-simplicial weakly self-dual factors, neither the
maximal nor minimal tensor product will be weakly-self-dual, since
$(A(\Omega) \otimes_{max} A(\Omega'))^* \cong A(\Omega)^*
\otimes_{min} A(\Omega')^* \cong A(\Omega) \otimes_{min} A(\Omega')
\ncong A(\Omega) \otimes_{max} A(\Omega')$.

If $A$ and $A'$ are order-isomorphic (alternatively, and notationally
easier, $A = A'$; hereafter this will be assumed), one candidate for
the weakly self-dual tensor product, namely, the convex hull of the
pure tensor states and the isomorphism states.  However, this is
unsatisfactory in various ways, not least that it degenerates to the
minimal tensor product if $A$ and $B$ are not isomorphic; also, using
this observation, it's easy to see that this tensor product is not
associative.

A better candidate is the {\em steering product}.

\begin{definition} Let $A \stensor B$, called the 
``steering product'' of $A$ and $B$, be $A \otimes B$, ordered by the
  cone generated by all steering states in $A \maxtensor B$.
\end{definition}

As remarked above, pure product states are steering; hence, this is a
valid tensor product in our sense.

\begin{conjecture} $\stensor$ is associative. \end{conjecture}

Suppose $\eta_{A} : A^{\ast} \rightarrow A$ and $\eta_{B} : B^{\ast}
\rightarrow B$ are order-isomorphisms implementing the weak
self-duality of $A$ and $B$. We can use these to convert a state
$\hat{\omega} : A^{\ast} \rightarrow B$ to an effect
\[\eta_{AB}(\omega) :=
\eta_{B}^{-1} \hat{\omega} \eta_{A} : A \rightarrow B^{\ast}\] This
gives us a linear map $\eta_{AB} : A \maxtensor B \rightarrow
A^{\ast} \maxtensor B^{\ast} = (A \mintensor B)^{\ast}$. Evidently,
this is positive. It is also easy to check that $\eta_{AB}$ takes
product states to product effects, and isomorphism states to
isomorphism effects.

\begin{question}
If $\hat{\omega}$ is steering, is $\eta_{AB}(\hat{\omega})$ also
steering (in some suitable dual sense)?
\end{question}

\begin{fact}
If $A = B = {\cal L}_h({\mathbb C}^{2})$---that is, if $A$ and $B$
are two qubits, then $A \stensor B = A \maxtensor B$ (which is not
weakly self-dual).
\end{fact}

\proof To see this one uses the fact that states in the maximal tensor
product all correspond to positive maps that are decomposable, i.e.,
sums of completely positive and co-completely positive maps.  The
extremal ones are all either product states, or isomorphism states,
since the automorphism group of a qubit---indeed, of any quantum
system---is generated by the maps $X \mapsto A X A^\dagger$ for
nonsingular $A$, and any transpose map $X \mapsto X^t$. \qed

One might ask whether if $A$ and $B$ are weakly self-dual, $A \stensor
B$ is too.  This is not so, as the above example of two qubits,
combined with observation that the maximal tensor product of
non-simplicial weakly self-dual cones is not weakly self-dual, show.

Same question if $A$ and $B$ are homogeneous.

\tempout{
\begin{question} Under what conditions can a state $\alpha \in A$ be
realized as the marginal of a steering state? What state spaces $A$
have the property that every state in $A$ is the marginal of a pure
state in $A \stensor A$? \end{question}

\begin{question} 
How does steering relate to teleportation? The idea of dual steering
is probably relevant here.
\end{question}
} 

\begin{question} 
Is the steering-for-both-marginals product equal to the
steering-for-one-marginal product?  This would follow, for example,
from the proposition that every steering-for-one-marginal state is a
convex combination of steering-for-both-marginals states.
\end{question}

\begin{question}
Is the steering-for-both-marginals product, perhaps under conditions
like homogeneity of both state spaces, equal to the topological
closure of the cone generated by automorphisms?
\end{question}

  We don't know the answer for sure even in the quantum case, because
  this raises the question whether states corresponding to
  nondecomposable maps can be steering.  We conjecture that they
  cannot, and that therefore the answer is ``yes'' in the quantum
  case.
  
\end{document}